\documentclass[preprint]{elsarticle}
\usepackage{amsmath,amssymb,latexsym}
\usepackage{amsfonts} \usepackage{psfrag}
\usepackage{graphicx}
\usepackage{epic}
\usepackage{eepic}
\usepackage{psfrag}
\usepackage[mathcal]{euscript}

\def\be{\begin{equation}}
\def\ee{\end{equation}}
\def\ba{\begin{eqnarray}}
\def\ea{\end{eqnarray}}

\def\lb{\label}
\def\H{{\cal H}}
\def\e{{\rm e}}
\def\f{\eta}

\begin{document}
\begin{frontmatter}

\title{Synchrotron radiation from massless charge}

\author{D. V. Gal'tsov}
\ead{galtsov@phys.msu.ru}

\address{Faculty of Physics, Moscow State University, Lebedev street 1, Moscow 119991, Russia}

\begin{abstract}
Classical radiation power from an accelerated massive charge
diverges in the zero-mass  limit, while some general arguments
suggest that strictly massless charge does not not radiate at all.
On the other hand, the regularized  classical radiation reaction
force, though looking odd, is non-zero and finite. To clarify this
controversy, we consider radiation problem in massless scalar
quantum electrodynamics in the external magnetic field. In this
framework, synchrotron radiation  is found to be non-zero, finite,
and essentially quantum. Its spectral distribution is calculated
using Schwinger's proper time technique for {\em ab initio} massless
particle of zero spin. Provided $E^2\gg eH$, the maximum in the
spectrum is shown to be at $\hbar \omega=E/3$, and the average
photon energy is $4E/9$. The normalized spectrum is universal,
depending neither on $E$ nor on $H$. Quantum nature of radiation
makes classical radiation reaction equation meaningless for massless
charge. Our results are consistent with the view (supported by the
renormalization group argument) that the correct classical limit of
massless quantum electrodynamics is free theory.
\end{abstract}


\end{frontmatter}

\section{Introduction}
Recently the problem of radiation from   massless charges attracted
attention in the framework of classical electrodynamics. It was
argued that the line singularity of the Lienard-Wiechert potential
for the massless charge  forbids using the Green  function method to
compute radiation \cite{Azzurli:2012tp,Azzurli:2014lha}, while the
conservation equation for Maxwell energy-momentum tensor implies
that the massless charge  does not  radiate   at all
\cite{Lechner:2014kua}. The similar assertion was earlier formulated
by Kosyakov \cite{Kosyakov:2007pm} basing on conformal invariance of
classical electrodynamics with massless charges (see also \cite{
Yaremko:2012zz}). On the other hand, a non-zero and finite
expression has been derived \cite{Kazinski:2002je,Lechner:2014kua}
for the regularized radiation reaction force acting upon an
accelerated massless charge. Finally, both these alternatives seem
to disagree with divergence of the classical formulas for radiation
power from an accelerated massive charge \cite{LL} in the limit of
zero mass. To clarify this confusion, one is led to consider the
radiation problem from quantum viewpoint.

   In quantum electrodynamics the limit of zero mass
also has  peculiar features. First, apart from  usual infrared
divergencies, there are {\em collinear singularities}, occurring
when photon is emitted from  massless legs in the Feynmann diagrams
in the direction of the  charge momentum \cite{Weinberg:1965nx}.
This is manifestation of degeneracy of states of the  charge and the
photon moving along the same line. Elimination of collinear
divergences is achieved using Kinoshita-Lee-Nauenberg
\cite{Kinoshita:1962ur,Lee:1964is,Nauenberg:1965uka} prescription to
 average over an ensemble of degenerate states (for review and further
references see \cite{Prokhorov:1999ri,Lavelle:2005bt}). Note that
the above mentioned line singularities of  classical retarded
potentials look as classical counterparts of collinear singularities
of quantum theory. Their regularization was proposed in
\cite{Lechner:2014kua} by imposing condition that the potentials
should be defined as distributions, the idea reminiscent of the
quantum averaging.

Second, vacuum polarization was shown to induce screening of the
massless charges created in high-energy processes
\cite{Fomin76,Gribov:1981jw,Morchio:1985re}.  However, this
screening occurs at very large distances of the order $L\sim
E^{-1}\exp(3\pi/\alpha)$ \cite{Smilga:1990uq}, where $\alpha$ is the
fine structure constant. It is therefore legitimate to consider
processes whose formation length is of the order of $E^{-1}$  (which
turns out to be our case). Anyway, though peculiarities of massless
QED  were often interpreted as manifestation of inconsistencies of
the theory or  proof  of non-existence of massless charges
 \cite{Fomin76,Gribov:1981jw}, the prevailing current opinion is that massless
QED is   special, but viable theory
\cite{Vaks:1961,Smilga:1990uq,Contopanogos:1991yc}. Moreover,
massless QED in the {\em external magnetic field} attracted much
attention during two past decades in connection with prediction of
the magnetic catalysis of the chiral symmetry breaking \cite{GMS},
 and other interesting phenomena relevant to
solid state physics and cosmology \cite{Ferrer:2012pb}. These
effects are associated with dynamics at   low Landau  levels in the
magnetic field. Here, exploring synchrotron radiation, we will deal
with high excitation levels. Note that, once the magnetic field is
treated non-perturbatively, the charges are never free,  so in this
approach there are no collinear singularities in the radiation
amplitudes.

 We will show that the theory predicts finite
synchrotron radiation from massless charges at the quantum level. We
will also investigate whether the zero mass limit of  the radiation
amplitudes obtained within the massive theory match with genuine
massless ones. We find that, unlike the situation in classical
theory, the massless limit for synchrotron radiation in quantum
theory is smooth in the quasiclassical (high Landau levels) regime.

\section{ Classical  theory with quantum cut-off}
Classical formula \cite{LL} for the total radiation power emitted by
a massive charge moving with the energy $E$ along the circle in the
magnetic field $H$,
 \be\lb{clas}
P_{\rm cl}=\frac{2e^4 H^2}{3m^2}\left(\frac{E}{m}\right)^2\,,
 \ee
diverges in the limit $m\to 0$. To clarify the origin of this
divergence let us pass to the spectral-angular distribution. The
famous Schott formula
 \be\lb{scht}
dP =\sum_{\nu=0}^{\infty} \frac{e^2\nu^2 \omega_H^2}{2\pi }\left[
 \cot^2\theta J_\nu^2(\nu \beta \cos\theta)+v^2 {J_\nu'}^2(\nu \beta \cos\theta)\right]
 d\Omega \,,\ee
where $E=m/\sqrt{1-\beta^2}$, shows that the frequency of radiation
is discrete and consists of the harmonics of the relativistic Larmor
frequency:
 \be
\omega= \nu \omega_H,\,\quad \omega_H=\frac{eH}{E}\,,
 \ee
and by virtue of the properties of Bessel functions,  the effective
domain of $\nu$ extends to
 \be \lb{nucr}\nu\lesssim\nu_{{\rm cr}}\sim (1-\beta^2)^{-3/2}= (E/m)^3
\,.
 \ee

The Schott formula itself does not depend on $m$ and admits the
limit of the velocity of light $\beta\to 1$. But  in the massless
theory there will be no frequency cutoff (\ref{nucr}), and this is
the reason why the total power (\ref{clas}) diverges in this case.
Therefore, for the massless charge, the cutoff frequency should
necessarily be quantum, $\hbar \omega_{\rm max}=E$, or $\nu_{\rm
max}=E^2/eH$ (assuming $eH>0$). This quantum cutoff does not depend
on the mass $m$ either and nicely fits with the quantization rule
for purely transverse motion \cite{Sokolov1}:
 $
E^2=E_n^2=m^2+eH(2n+1)\,,  n=0,1,2,\ldots  $ , provided both initial
and final level numbers  $n,\; n'$ are large and $\nu=n-n'\ll n\,.$

Integrating (\ref{scht}) over angles  and passing to continuous
frequency distribution $P=\int P(\omega)\,d\omega$, one can show
that the low frequency limit of $P(\omega)$ is also
mass-independent:
 \be
P_{\rm low }(\omega)=\frac{e^2 \omega_{H}3^{1/6}\Gamma(2/3)}{ \pi }
\left(\frac{\omega}{\omega_H}\right)^{1/3} \,,
  \ee
so we can integrate it up to $ \omega_{\rm max}=E/\hbar$ to get an
estimate for the total radiation power:
 \be \lb{cut} P_{\rm cut} =\frac{e^2 \sqrt{3} \Gamma(2/3)}{4\pi \hbar^2} (3e\hbar HE)^{2/3}
 \,.
 \ee
This quantity is finite, but  contains a factor $\hbar^{-4/3}$,
showing that radiation is essentially quantum.
\section{Mass operator for $m=0$}
Quantum theory of synchrotron radiation has different formulations.
Historically, the first was a direct approach making use of exact
solutions of the wave equations in magnetic field \cite{Sokolov1}
(for a more recent review see \cite{Bor}). Later on Schwinger
suggested the ``proper time'' method to calculate the mass operator
of the charge in the constant field $F_{\mu\nu}$ for spins
$s=0,\,1/2$; its imaginary part  gives the total probability rate of
synchrotron radiation \cite{Schwinger:1973,Tsai:1974cc}. Further
ramifications of this approach allow to get spectral and
spectral-angular distributions of radiation \cite{Schwinger:1977ba}
(for alternative constructions of the mass operator in constant
electromagnetic field see \cite{Baier:1974qq,Ritus:1978cj}). Here we
apply the technique of \cite{Schwinger:1973,Tsai:1974cc} for
zero-spin charged particle of strictly zero mass. The massless limit
of the corresponding massive theory will be discussed in the next
section.

   The action term involving the mass operator for the complex scalar
field $\phi(x)$ is
 \be\lb{first}
  -\frac12\int\phi(x)M(x,x')\phi(x')dx dx',
\ee where $M(x,x')$  in Schwinger's operator notation reads:
 \be
 M=ie^2\int\left[(\Pi-k)^\mu\frac1{k^2}\frac1{(\Pi-k)^2}(\Pi-k)_\mu\right]\;
 \frac{d k}{(2\pi)^4}\;
 \;-M_0\,.
 \ee
Here $
 \Pi_\mu=-i\partial_\mu-e A_\mu,\;\,A_\mu
$ denotes the constant magnetic filed, and $M_0$ is the subtraction
term needed to ensure vanishing of $M$ and its first derivative with
respect to $ \Pi^2$ at $\Pi^2=0$. It consists of the first two
Taylor expansion terms of $M$ in $ \Pi^2$ for $F=dA=0$. The idea of
the proper time method is to perform
 exponentiation of  propagators
 \be
\frac1{k^2}\frac1{(\Pi-k)^2 }=-\int_0^\infty
sds\int_0^1\e^{-is\H}\,,\quad \H=(k-u\Pi)^2-u(1-u) \Pi^2\,,
 \ee
and to replace the $k$-integration by averaging over the eigenstates
of the operator $\xi_\mu$, canonically conjugate to $k_\mu$,
$[k_\mu, \xi_\nu]=i\eta_{\mu\nu}$ in the Hilbert space of a
fictitious particle:
 \be
M=ie^2\int_0^\infty sds\int_0^1 du   \langle
\xi\big|(\Pi-k)^\mu\e^{-is\H} (\Pi-k)_\mu\big|\xi\rangle \;\; -M_0
\,.
 \ee
The quantity $\H$ is then treated as Hamiltonian of this particle
and the averaging is performed in the Heisenberg picture passing to
$s$-dependent operators $k(s), \xi(s), \Pi(s)$, which can be found
exactly in terms of $F$. Performing calculations and comparing the
results with formulas given in \cite{Tsai:1974cc} for the massive
theory we find that one can start with the Eq. (40) of Tsai
\cite{Tsai:1974cc} setting there $m=0$. The subsequent computations
in \cite{Tsai:1974cc} use approximations of the integrals in terms
of the Macdonald functions. These approximations become singular as
$m\to 0$, so we develop an alternative integration scheme.

As in \cite{Tsai:1974cc}, our starting formula is valid  for the
{\em on shell} mass operator, i.e. for
$\phi(x)=\phi({\bf{r}})\e^{-iEt}$ satisfying the wave equation
$\Pi^2 \phi=0$, with discrete eigenvalues of the energy (we consider
purely transverse motion) \be E=\sqrt{eH(2 n+1)}\,. \ee It reads:
 \be\lb{Mint}
M=\frac{e^2}{4\pi}\int_0^1 du\int_0^\infty
\frac{ds}{s}\left[\e^{-i\psi}\Delta^{-1/2}\left(E^2\Phi_1+4ieH\Phi_2+
 i \Phi_3/s\right)-2i/s\right]\,,
  \ee with
\begin{align} \lb{Phis}
&\Phi_1=3-4u+u^2-\frac{(1-u)^2}{\Delta}(4 \cos 2x
-1)-\frac{u(1-u)}{x\,\Delta}\sin 2x\,,\nonumber\\
 & \Phi_2=\sin 2x-\frac{2u(1-u)\sin^2 x}{x\,\Delta}\cos 2x\,,\nonumber\\
 & \Phi_3=1+\frac{1-u}{\Delta}(2\cos 2x -1)+\frac{u\sin
 2x}{2x\,\Delta}(4\cos 2x -3)\,,\end{align} where $x=eHsu$,   and
  \begin{align}
& \Delta=(1-u)^2+u(1-u)\frac{\sin 2x}{x}+u^2\left(\frac{\sin
 x}{x}\right)^2\,,\nonumber\\
& \psi=(2n+1)[\beta-(1-u)x]\,,\quad \tan\beta= \left(\cot
x+\frac{u}{x(1-u)} \right)^{-1}\,.\nonumber
\end{align}
This expression is valid for all Landau levels $n$.

In what follows we will be interested in the case $n\gg 1$, when the
integrals can be evaluated expanding all $x$-dependent quantities in
power series. Indeed, the main contribution to the integral over $x$
comes from the region where the  phase $\psi(x,u)\lesssim 1$, in
which $\beta$ for $x\ll 1$ can be approximated as
 \be
\beta \approx (1-u)x+u(1-u)^2x^3/3\,,
  \ee
so that
 \be \lb{psix}
\psi\approx(2n+1)\alpha x^3=\frac{E^2}{eH}\alpha
x^3=\frac{s^3}3(eHE)^2 u^4(1-u)^2 \,,\quad \alpha=u(1-u)^2/3\,.
 \ee
For large $n$, apart form the narrow regions around the limiting
points of $u$,
 \be
u>n^{-1}\,,\quad 1-u>n^{-1/2},\,
 \ee
 which {\em a posteriori} give negligible contributions, the essential
 domain of $x$ is
 \be \lb{xbound}
  x\lesssim  n^{-1/3}\,.
 \ee
Therefore we use (\ref{psix}) in the exponent, expanding the other
functions in powers of $x$ \footnote{The initial integral
(\ref{Mint}, \ref{Phis})  converges at the upper limit $x\to
\infty$, while  higher terms in Taylor expansion of the integrand
will produce divergent quantities, what is typical for asymptotic
series. We will keep only the leading terms giving the convergent
integrals over $x$.}:
 \begin{align}
&\Delta^{-1}\approx 1+u(4-3x) x^2/3\,, \\\lb{fi1x}
&\Phi_1\approx \left(8- {32u}/{3} +13 u^2/3-  u^3 \right)(1-u)x^2\,,\\
&\Phi_2\approx 2(1-u+u^2)x\,,\\
&\Phi_3\approx 2-(4- {10 u}/{3}+u^2)x^2\,.
  \end{align}
Using the bound (\ref{xbound}) and taking into account different
orders  of various terms in $n$, we find that the contribution of
$\Phi_2$ will be of the order of $n^{-1/3}$ with respect to the
leading term $\Phi_1$, while in $\Phi_3$ one has to keep only the
zero order term. Introducing the decay rate via
 \be
\Gamma=-\frac1{E}\;{\rm{Im}}\, M\,,
 \ee
 we obtain
 \be
 \Gamma=\frac{e^2}{4\pi E}\int_0^1 du \int_0^\infty
 \frac{dx}{x}\left(E^2\Phi_1\sin\psi\;
 +2\frac{eHu}{x}(1-\cos\psi)\right)\,,
\ee where for $\psi$ one has to use (\ref{psix}), and for $\Phi_1$
--- the approximation (\ref{fi1x}). Taking into account the table
integrals
\begin{align}
 & \int_0^\infty \sin (z^3)\, z
  dz=\frac{\Gamma\left(2/3\right)}{2\sqrt{3}}\,,\\
 & \int_0^\infty \left[1-\cos (z^3)\right]\frac{dz}{z^2}=\frac{\sqrt{3}\Gamma\left( 2/3\right)}2
  \,,
\end{align}
we get
 \be
 \Gamma=\frac{e^2 \Gamma\left( 2/3\right)\left(3eHE\right)^{2/3} }{8\pi\sqrt{3}E}
 \int_0^1  \frac{8- {32u}/{3} +19 u^2/3- 3 u^3}{u^{2/3}(1-u)^{1/3}}du\,.
 \ee
The integrand has an integrable singularity at the lower limit.
Integrating, we finally obtain the total decay rate
 \be\lb{Ga}
 \Gamma=\frac{4e^2}{9E}\Gamma\left( 2/3\right)\left(3eHE\right)^{2/3}\,.
 \ee
\section{The spectral power }
To get the spectral power of radiation one has to perform  Fourier
expansion  in the mass operator, leading to
 \be
{\rm{Im}} M= {\rm{Im}}\left( \int d\omega\int_{-\infty}^{\infty}
\e^{i\omega \tau} \,M' \;\frac{d\tau}{2\pi}\right)\,,
 \ee
where
 \be
M'=-ie^2\int_0^\infty sds\int_0^1 du   \langle
\xi\big|(\Pi-k)^\mu\e^{-is\H}\e^{-ik^0\tau}
(\Pi-k)_\mu\big|\xi\rangle -M'_0 \,.
 \ee
The spectral power $P(\omega)$ is then introduced via the relation
 \be
\Gamma=\int P(\omega)\frac{d\omega}{\omega}\,,
 \ee
 so one obtains:
 \be
P(\omega)=-\frac{\omega}{E}\;{\rm{Im}}\left(\int_{-\infty}^{\infty}
\e^{i\omega \tau} \,M' \;\frac{d\tau}{2\pi}\right)\,.
 \ee

\begin{figure}
\begin{center}
\includegraphics[angle=270,width=8cm]{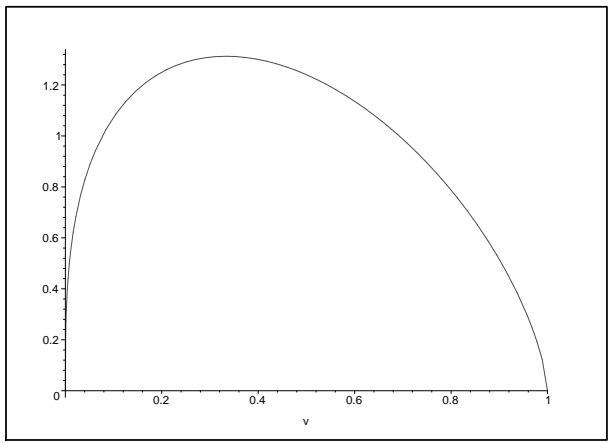}
\caption{\small The universal normalized spectral distribution
${\cal P}\left(v\right)$ (\ref{speless}) of synchrotron radiation
from massless scalar charge with the energy $E$ satisfying $E^2\gg
eH$ versus $v=\hbar\omega/E$. The curve has maximum at $v=1/3$, the
average photon energy ${\hbar \omega}=4E/9$.}\label{F1}
\end{center}
\end{figure}

With this modification we find:
  \begin{align}
\!\!\!\!P(\omega)= -\frac{e^2}{4\pi}\frac{\omega}{E}\;{\rm{Im}}
 \int_{0}^{\infty} \frac{ds}{s}\int_0^1 du &
  \Bigg[\frac{E^2x^2}3e^{-i\psi}\left(24-  56u  + 45 u^2 -  16u^3+3u^4\right) + \nonumber\\
& +\left(e^{-i\psi}-1 \right)\left( \frac{2i }{s} - \frac{ i(2-u)
}{s}\frac{d}{du}  \right)\Bigg] J \,,
  \end{align}
where the non-leading terms were omitted, and $J$ denotes the
  integral over $\tau$:
  \be
 J(\omega,
u)=\frac1{2\pi}\int_{-\infty}^{\infty}\e^{i(\omega-uE)\tau-i\tau^2/4s}
d\tau\,.
  \ee
For large $E$, this integral is formed in the region $uE\tau\lesssim
1$, where the quadratic factor in the exponent has the order
 \be
\frac{\tau^2}{4s}\sim \frac{eH}{4uE^2x}=\frac1{4(2n+1)ux}\,.
  \ee
For $n\gg 1$, the effective domain in the $x$-integral is $x\lesssim
n^{-1/3}$, thus the whole factor has the order $n^{-2/3}$. Omitting
this term in the exponent, we  obtain  the delta-function
  \be
J(\omega, u)\approx\delta(\omega-uE)\,,
  \ee
which is then used to integrate over $u$. The derivative term is
evaluated by parts. Differentiating the phase $\psi$, one has to
take into account that the independent variables are $(s,u)$, so one
has to use for $\psi$ the last expression in (\ref{psix}) giving
 \be
\partial_u \psi=\frac{2E^2x^3}{3eH}(2-3u)(1-u)\,.
  \ee
Then, denoting $v=\omega/E$, we obtain
 \be
P(\omega)= \frac{e^2v}{4\pi E}\int_0^\infty
  \left(E^2(8-v^2)(1-v)^2x\sin\psi\;
 +\frac{eHv}{x^2}(1-\cos\psi)\right)dx\,,
 \ee
  The  integral over $x$ is
evaluated as before, and finally  we get
 \be\lb{speless}
P(\omega)=\frac{2e^2\;\Gamma\left(2/3\right)}{27\hbar
E}\,\left(3e\hbar HE\right)^{2/3}{\cal P}\left(
{\hbar\omega}/{E}\right)\,,
 \ee
 where the Planck's constant is restored, and the
normalized spectral function is introduced \be \lb{spenorm} {\cal
P}\left(v\right)=\frac{27}{2\pi\sqrt{3}}\;v^{1/3}(1-v)^{2/3}\,,\qquad
\int_0^1{\cal P}\left(v\right)dv=1\,.
 \ee
This spectrum, shown in Fig.~\ref{F1}, is  perfectly smooth
function, whose maximum lies at \be \hbar \omega_{\rm max}=\frac13
E\,. \ee The average   photon energy is
 \be
\langle\hbar\omega\rangle=E\int_0^1  v{\cal P}\left(v\right)dv=
\frac49 E \,.
 \ee
 The total
energy loss per unit time,
 \be\lb{tot}
P=\int_0^{E/\hbar} P(\omega)d\omega=
\frac{2e^2\;\Gamma\left(2/3\right)}{27\hbar^2 }\;\left(3e\hbar
HE\right)^{2/3}\,,
 \ee
differs from the estimate (\ref{cut}) only by a numerical
coefficient.

Synchrotron radiation of the massless charge is therefore purely
quantum, its intensity being divergent when $\hbar\to 0$. It
consists of energetic quanta of the order of the charge energy. The
spectral distribution of radiation in the leading order in inverse
quantum number $1/n$ is given by a universal formula which does not
depend on the magnetic field at all. So even in a weak magnetic
field the massless charge converts its energy into quanta of the
same order of energy.
\section{Massless limit of the massive theory}
Transition to the massless limit in the quantum theory of
synchrotron radiation of massive charge  is subtle, since the
results of the latter depend on two dimensionless parameters (in
units $\hbar=c=1$):
 \be\lb{param}
\f=\frac{H}{H_0}=\frac{eH}{m^2}\,,\qquad \chi= \frac{H}{H_0}
\frac{E}{m}=\frac{eHE}{m^3}\,,
  \ee
which both  diverge as $m\to 0$. This is why we started by
considering radiation in the massless theory {\em ab initio}. Now we
are going to discuss the massless limit in some results of the
massive theory which are known analytically.

Simple closed formulas for the spectral power and the total
intensity of synchrotron radiation from scalar charge exist for
$\f\ll 1$ and arbitrary $\chi$ \cite{Matveev,Bagrov65}. This case
corresponds to transitions between Landau levels $n\gg 1$ and $n'\gg
1$  giving the dominant contribution in this case. Sub-expansions in
terms of $\f$ were explored in
 \cite{Sokolov:1973zv,Baier:1990qq,DBZ}. It was shown that for large  $\chi$
the leading  term remains dominant provided $\chi\gg \f$ even if
$\f$ itself is not very small. Note, that the two parameters
(\ref{param}) have different orders in $1/m$, so that the ratio $
{\f}/{\chi} = {m}/{E} $ tends to zero in the massless limit.
Therefore, one can hope to get sensible results using formulas
obtained in the case $\f\ll 1$.

The total probability and the total power of radiation in the
massive $\f\ll 1$ case are known to be mass-independent in the
so-called ultra-quantum limit $\chi\gg 1$ \cite{DBZ} \footnote{I am
indebted to A.~V.~Borisov for drawing my attention to this fact.}.
Closer look reveals that they are given {\em precisely} by our
formulas (\ref{Ga}) and (\ref{tot}) obtained in the strictly
massless theory. Thus we have proven that even in the case $\f\to
\infty$ the leading term in $\chi$ of the massive theory remains
untouched if $\chi/\f\to \infty$ as well. Moreover, the {\em
spectral distribution} of synchrotron radiation in the massive
quantum theory obtained for $\f\ll 1$ also tends to our universal
power spectrum (\ref{spenorm}).  Indeed, the spectral distribution
found in \cite{Matveev,Bagrov65,Tsai:1974cc} for $\f\ll 1$ and
arbitrary $\chi$ in our notation reads:
 \be \lb{spemas}P_{\rm mass }(\omega)=\frac{ e^2m^2\omega}{\sqrt{3}
E^2}\int_y^\infty K_{5/3}(\xi) d\xi\,,\quad
y=\frac{2\omega}{3(E-\omega)\chi}\,.
 \ee where $K_{5/3}$ is the Macdonald function.
In the limit $m\to 0$ one has $\chi\to \infty$ and $y\to 0$, and,
using l'Hopital rule
 \be
\int_y^\infty K_{5/3}(\xi) d\xi \sim 2^{2/3} \Gamma\left(2/3\right)
y^{-2/3}\,,
 \ee
we see that the spectral power (\ref{spemas}) reduces to
(\ref{speless}, \ref{spenorm}) indeed. Thus our calculations in the
preceding section prove that the standard quasiclassical
approximation derived in the massive theory for $\f\ll 1$ and any
$\chi$ remains valid in the ultra-quantum limit $\chi\gg 1$ not only
for  finite $\f$, but also for $\f\to\infty$, provided $\chi/\f\to
\infty$ as well.

Similarly, using the results of \cite{Bagrov65}, one can  perform
transition to zero mass in the amplitudes for emission of photons
with different polarizations. One finds that the linear polarization
in the plane orthogonal to magnetic field $P_{\sigma}$ prevails
above the orthogonal component $P_{\pi}$:
 \be
                    {P_{\sigma}}=3{P_{\pi}} \,.
  \ee

\section{Magnetically induced $m^2$}
   In scalar massless QED it is natural to
consider magnetic generation of the {\em square} of mass $m^2$,
since just this quantity enters the Klein-Gordon equation. In the
definition  (\ref{first}) the mass operator has dimensionality of
$m^2$, so its real part gives correction $\delta  m^2 $. This
quantity is finite for $m=0$. Had we extracted  (as in
\cite{Ritus:1978cj,Baier:1990qq})  the linear quantity $\delta m$
instead, we would get an infinite result: since $\delta  m^2 =2
m\,\delta m$, the linear correction $\delta m=\delta  m^2/(2m)$
diverges  as $m\to 0$. Thus $\delta m$ is meaningless in the
massless case.

Keeping only the leading terms in the real part of (\ref{Mint})  we
find:
 \be
\delta m^2={\rm Re }M=\frac{e^2}{4\pi }\int_0^1 du \int_0^\infty
 \frac{dx}{x}\left(\cos\psi\;
 \Phi_1+2 \sin\psi\frac{eHu}{x}\right)\,.
  \ee
Using the integrals
\begin{align}
 & \int_0^\infty \cos(z^3)\, z
  dz=\frac16 \Gamma\left(2/3\right) \,,\\
 & \int_0^\infty  \sin (z^3) \frac{dz}{z^2}=\frac12 \Gamma\left( 2/3\right)
  \,,
\end{align}
we obtain:
 \begin{align}
  \delta m^2=  \frac{e^2}{4\pi }\int_0^1
\frac{8-32u/3+19u^2/3-3u^3}{u^{2/3}(1-u)^{1/3}}du=
 \frac{4e^2\;\Gamma\left(2/3\right)}{9\sqrt{3}}\;\left(3e
HE\right)^{2/3} \,.
 \end{align}
This expression, non-perturbative in $H$,  is valid for any $H$,
provided $E^2\gg eH$.
\section{Stochastic nature of radiation reaction force}
Recently  radiation reaction problem for  massless charge attracted
some attention. In \cite{Kazinski:2002je} the closed formula for
classical  reaction  force was derived including  three divergent
terms.  Comparing these results with those known in the massive
case, one notices several strange features. First, there is no
intrinsic parameters which could absorb divergencies, like mass in
the massive theory, while the number of divergent terms is increased
from one to three. Second, there are divergent terms of
non-lagrangian nature which can not be incorporated into the action.
Third, the finite part of the reaction force contains quite a high
derivative (fifth), and no examples are known to compare it with the
radiation power. Finally, the   reaction force can not be obtained
from the finite Dirac-Lorentz force known in the massive case since
the latter diverges in the limit of zero mass.

The formal result of \cite{Kazinski:2002je} was confirmed in
\cite{Lechner:2014kua} using different regularization, but the
conclusion of \cite{Lechner:2014kua} was that the reaction force has
no physical meaning, since massless charge does not radiate at all.
As we have shown here,  the massless charge does radiate and the
radiation power is finite. However, it is essentially quantum and
the average energy of the photons is of the order of the particle
energy. The reaction force is therefore  stochastic, and
consequently,  the classical Lorenz-Dirac type equation derived in
\cite{Kazinski:2002je} is meaningless indeed.

Moreover, although the stochastic nature of the radiation recoil is
already enough to preclude any classical radiation reaction
equation, in the theory of synchrotron radiation of massive charges
there also exists a much stronger restriction on the validity of
such an equation, known as the $E_{1/5}$ bound \cite{Sokolov1}. The
reason is that the quantum states of the charge in the magnetic
field for the transverse motion depend on {\em two} quantum numbers,
the second one being responsible for location of the center of the
orbit \cite{Sokolov1}. Quantum fluctuations due to excitation of the
orbit center  start at much lower energies than quantum recoil comes
into play. Namely, for $\f\ll 1$,
 \be
E_{\rm recoil}\sim m \frac{H_0}{H}\,,\qquad E_{\rm fluct}\sim
E_{1/5}= m \left(\frac{H_0}{H}\right)^{1/4}\,,
 \ee
so $E_{\rm fluct}\ll E_{\rm recoil}$. Classical description of
radiation reaction is valid for $E\ll E_{1/5}$. Now, for massless
particles $E_{1/5}=0$, so classical radiation reaction is twice
meaningless.

\section{Discussion}
We have shown that quantum scalar electrodynamics in the external
magnetic field unambiguously predicts finite synchrotron radiation
in the form of emission of hard photons with energy  of the order of
the particle energy. We have found this  first using the strictly
massless theory, and then considering the zero-mass limit of the
massive theory. Both give the same results for the total
probability, the total radiation power and the spectral distribution
of radiation, provided $E^2\gg eH$, i.e. for high Landau excitation
levels. This is in sharp contrast with the situation in classical
theory, where there is no smooth transition from massive to massless
theory: radiation should be absent in massless theory in view of
conformal invariance \cite{Kosyakov:2007pm} and conservation
equations \cite{Lechner:2014kua}, though the radiation power of the
massive theory diverges in the $m\to 0$ limit. These two features do
not contradict each other, however. Indeed, it was argued long ago
that in the massless theory the renormalization group flow has
stable infrared point corresponding to vanishing charge \cite{BSH}.
So, from the point of view of the quantum theory, correct massless
classical electrodynamics must be  non-interacting.

The normalized spectrum of photons emitted by the massless charge in
the magnetic field does not depend on any parameter. Thus, even in a
very small magnetic field like that of the Earth, massless charges
of  the energy satisfying  $E^2\gg eH$ will emit photons with
energies of the same order. Also, since we have shown that the limit
to zero mass in the massive theory is smooth,   our universal
spectrum (\ref{spenorm}) has not only academic interest, but also
applies to   massive particles with energies $E/m\gg \eta^{-1}$ {\em
for any} $\f$.

Another interesting prediction following from our calculations   is
magnetic generation of the square of mass in the linear order in
$\alpha$. This was obtained taking the real part of the mass
operator, and also can be derived via dispersion relation from the
synchrotron radiation rate.

We have considered here only the quasiclassical case $n\gg 1$.  Note
that in Schwinger approach summation over the finite quantum numbers
$n'$ is performed automatically, so in the leading in $1/n$
approximation our results are not restricted by the condition
$\nu=n-n'\ll n$. For low initial Landau levels $n$ the problem must
be treated numerically. Partially, the results may be extracted from
the existing numerical calculations performed in the massive theory
(see \cite{DBZ} and references therein).

\section*{Acknowledgements}
The author is grateful to  V.~G.~Bagrov, V.~A.~Bordovitsyn,
V.~R.~Khalilov and V.~Ch.~Zhukovsky for useful conversations, and
especially to A.~V.~Borisov for discussing some subtleties of the
Schwinger method. This work was supported by the RFBR grant
14-02-01092.

\end{document}